\documentclass{elsart}
\usepackage{graphicx}
\usepackage{amscd}
\journal{Nuclear Instruments and Methods}

\newcommand{\grb}{$\gamma$-ray burst}
\newcommand{\wxm}{{\em WXM}}

\begin{document}
\input epsf

\begin{frontmatter}
\title{A Fourier-Based Algorithm for Modelling Aberrations in HETE-2's
Imaging System}
\author[riken]{B. M. Sch{\"a}fer\thanksref{now}\corauthref{cor}},
\corauth[cor]{corresponding author. 
phone: +49 (0)89 30000 2272, fax: +49 (0)89 30000 2235}
\thanks[now]{Present address: Max-Planck-Institute for Astrophysics,
Karl-Schwarzschild-Stra{\ss}e 1, 85741 Garching, Germany}
\ead{spirou@mpa-garching.mpg.de}
\author[riken,daigaku]{N. Kawai}
\ead{nkawai@hp.phys.titech.ac.jp}

\address[riken]{
RIKEN (The Institute of Physical and Chemical Research), 
2-1 Hirosawa, Wako, Saitama 351-0198, Japan}
\address[daigaku]{
Department of Physics, Faculty of Science, Tokyo Institute of
Technology, \mbox{2-12-1 Ookayama}, Meguro-ku, Tokyo 152-8551, Japan}

\begin{abstract} 
The High-Energy Transient Explorer (HETE-2), launched in October 2000, is a 
satellite experiment dedicated to the study of $\gamma$-ray bursts in a very 
wide energy range from soft X-ray to $\gamma$-ray wavelengths. The 
intermediate X-ray range (2-30~keV) is covered by the Wide-field X-ray 
Monitor {\wxm}, a coded aperture imager. In this article, an algorithm for 
reconstructing the positions of {\grb}s is described, which is capable of 
correcting systematic aberrations to approximately $1'$ throughout the field 
of view. Functionality and performance of this algorithm have been validated 
using data from Monte Carlo simulations as well as from astrometric 
observations of the X-ray source {\em Scorpius X-1}.
\end{abstract}

\end{frontmatter}

\section*{Introduction}

Although the number of detected $\gamma$-ray bursts has increased tremendously
due to the BATSE instrument on board the Compton $\gamma$-Ray Observatory
\cite{batse}, unique identifications and quick follow-up observations of the
celestial object that harbours the burst site are rare. This issue has been
successfully addressed by the BeppoSAX satellite, that monitors a large
fraction of the sky ($\sim 1\mbox{ sr}$) in the intermediate X-ray range 
\cite{bepposax1,bepposax2} and is able to derive burst localisations with an 
accuracy of $\sim 1'$. With such improved localisations, the Beppo-SAX team 
has succeeded in detecting X-ray afterglows \cite{glow}, by which valuable 
insight into the emission mechanism was gained. Finding $\gamma$-ray burst 
counterparts and performing spectroscopy at wavelengths other than the 
$\gamma$-ray band during the active phase of a burst as well as providing 
good localisations for follow-up observations is the basic scientific motivation 
of the HETE-2 mission.\newline

This publication is structured as follows: After a brief description of HETE-2's
instrumentation in section \ref{cont:instruments}, the position reconstruction
algorithm is outlined in section \ref{cont:algorithm}. The performance of the 
algorithm on Monte Carlo generated events as well as on data taken in 
astrometric observations of the X-ray source {\em Scorpius X-1} is 
presented in section \ref{cont:result}. In section \ref{cont:psf} findings on 
the point spread function are shown. A summary in section \ref{cont:conclusion} 
concludes the paper.

\section{HETE-2 Mission\label{cont:instruments}}
The High Energy Transient Explorer (HETE-2) is a dedicated mission for the
localisation and spectroscopy of $\gamma$-ray bursts. Details of the 
instrumentation of HETE-2 and the mission are given in \cite{hete}. Its 
scientific payload consists of three experiments: {\em FREGATE}, a 
scintillation crystal experiment, that is expected to provide triggers because of 
its high sensitivity and large sky coverage, {\em SXC}, a coded-mask imager 
based on a X-ray CCD chip design, which is able to localise $\gamma$-ray 
bursts with very high spatial resolution, and the core experiment {\wxm}. 
\newline

The Wide Field X-ray Monitor {\wxm} consists of two perpendicularly oriented
1-dimensional coded mask cameras, in which photons are detected by position
sensitive proportional counters.
The detectors, one pair for each of the two orthogonal systems, are filled with
xenon gas at a pressure of $1.4\mbox{ bar}$ with an admixture of $3\%$ carbon
dioxide as a quenching gas. Each detector contains three carbon wire anodes in which
the position of an absorbed photon is inferred by comparing the accumulated charges 
at both ends of a wire. {\wxm} achieves a position resolution of $1\mbox{ mm}$ 
(FWHM) at $6\mbox{ keV}$. A veto system below the counting wires reduces the 
background due to charged particles. {\wxm} is sensitive to photons with energies 
$E$ in the range $2\mbox{ keV}\leq E\leq 30\mbox{ keV}$ with an relative energy 
resolution of $\Delta E/E=8\%$ at $E=8\mbox{ keV}$.\newline

\begin{figure}[htb]
\begin{center}
	\mbox{
	\epsfxsize10cm
	\epsffile{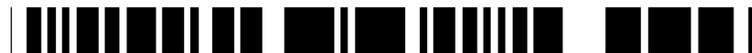}}
\end{center}
\caption{The coded mask of HETE-2's wide-field X-ray monitor {\wxm}}
\label{fig:hete}
\end{figure}

The mask pattern, identical for the $x$- and $y$-system, is placed
$d=187\mbox{ mm}$ above the detectors and is composed of 103 elements, each
$2\mbox{ mm}$ in width. On third of the elements are open and randomly
distributed. The parameters have been optimised with respect to source
localisation accuracy for the expected levels of signal and background photon
count rates \cite{zand}. The combined $x$- and $y$-systems are monitoring a
field of view of approximately $60^\circ\times 60^\circ$. Figure \ref{fig:hete}
shows the actual mask pattern of a {\wxm} camera. Further details of the {\wxm}
cameras can be found in \cite{yuji}.

In order to provide other experiments with precise localisations of $\gamma$-ray
burst sites within seconds after burst onset, an array of twelve burst alert stations
has been installed below HETE-2's flight path. Once received, the information is
relayed to the GRB Coordinate Network (GCN), from where it may be obtained by
interested observers.

\section{Position Reconstruction Algorithm\label{cont:algorithm}}
\begin{figure}[htb]
\begin{center}
\mbox{
  \epsfxsize5.5cm
        \epsffile{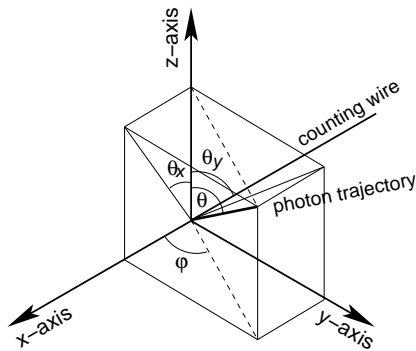}}
\end{center}
\caption{The definition of geometrical quantities in the {\wxm} coordinate system.}
\label{fig:angles}
\end{figure}

For the sake of readability, the algorithm is described for determining the 
angle of incidence $\theta_x$ in the $x$-detector, completely analogous 
formulae apply to the $y$-system. Figure \ref{fig:angles} provides the
definition of all quantities involved.

\subsection{Correlation}
The task of localising a single point source is performed by computing the
correlation function $\left[f\otimes m\right](x)$ of the mask pattern $m(x)$ and
the recorded intensity distribution $f(x)$; the asterisk denoting complex
conjugation:

\begin{equation}
\left[f\otimes m\right](x)=\int m(\lambda)\cdot f^*(\lambda+x) d\lambda\mbox{.}
\label{eqn:correlation}
\end{equation}
The correlation function $[f\otimes m](x)$ will peak
at a value $x_0$, indicating the distance $f(x)$ is shifted with respect to $m(x)$. 
This yields the angle of incidence $\theta_x$ by $\tan\theta_x=x_0/d$, where
$d=187\mbox{ mm}$ is the distance between mask and detector. In $m(x)$,
open elements of the mask pattern have been assigned a value of $1$, whereas
closed elements correspond to a value of $\tau/(\tau-1)=-1/2$, where $\tau=1/3$ is
the open fraction of HETE-2's imaging system. This method is known as balanced
correlation and removes the correlation background.

There exist more sophisticated correlation schemes such as unbiased balanced correlation
(e.g. \cite{ponman}, \cite{andrew}), that do take into account partial 
shadowing of the mask onto the detector by bursts in the periphery of the field of view and 
increased coding noise caused by steady strong off-axis sources. For HETE-2's flight algorithm
we rely on basic balanced correlation for three reasons: The width of the mask is almost twice as 
large as the active area of the detector and only bursts at angles $\geq 25^\circ$ off the optical 
axis are affected by partial shadowing. Furthermore, the running average of the mask's 
transparency is close to constant with deviations being very small, so that the 
modulation of the pattern is insensitive to angle of incidence. Thirdly, in HETE-2's nominal 
survey mode, the satellite is rotated in such a way that strong X-ray sources are avoided and the 
assumption of isotropically incidenting background radiation is valid.

\subsection{Imaging Aberrations}
In reality the correlation is complicated by imperfections of the detector: The
photons travel a finite path before interacting and the loci of photons
interacting in the detector are measured with finite position resolution (in
the case of {\wxm} to $1\mbox{ mm}$, subtending an angle of $18'$). 
Whereas the finite position resolution simply results in a blurring of the image, the 
penetration of energetic photons effectively induces an additional shift in the image, 
if the event is happening at nonzero angles of incidence, and causes the angles of
incidence to be overestimated. This effect increases in a complex fashion
with distance from the center of the field of view and is of the order of
$12'$ at $30^\circ$ from the optical axis for a photon distribution
following a power law spectrum with a spectral index of $\Gamma=1.1$.

In contrast to a time consuming Monte Carlo simulation, in which one
determines the image of the mask pattern under estimated angles of incidence,
the algorithm presented in this article aims at modelling all effects involved
in the delapidation of the mask pattern, enabling a much faster localisation of
the $\gamma$-ray burst site.

\subsection{Convolutions}
Imaging aberrations may be described by applying suitable changes to the mask 
pattern $m(x)$ prior to the calculation of the correlation function. This can 
be archieved by convolving the mask pattern $m(x)$ with integration 
kernels $p(x)$ and $r(x)$, that describe the penetration of highly energetic 
photons and the detector resolution, respectively,

\begin{equation}
\left[p\odot m\right](x)=\int m(\lambda)\cdot p(x-\lambda) d\lambda\mbox{.}
\label{eqn:convolution}
\end{equation}

The convolution kernel $p(x)$ that is used for describing the penetration of
photons is given by equation \ref{eqn:penetrate}:

\begin{equation}
p(x) = \frac{1}{a_{\|}(\theta_x,\theta_y)}\exp\left(-\frac{x}{a_{\|}(\theta_x,\theta_y)}\right)
\cdot\Theta\left(\mbox{\em sgn}\left[\theta_x\right]\cdot x\right)\mbox{,}
\label{eqn:penetrate}
\end{equation}

where $\Theta$ is the Heaviside function and {\em sgn} the signum function 
The convolution yields an altered mask pattern $m'(x)=[p\odot m](x)$, in which the 
penetration of photons is incorporated.
The length scale $a_{\|}$ of the exponential decay, given by equation \ref{eqn:ceqn}, is 
equal to the spectrally averaged attenuation length $a$ of photons inside the 
detector, corrected by a projection factor:

\begin{equation}
a_{\|}(\theta_x,\theta_y)= 
a\cdot\frac{\tan\theta_x}{\sqrt{1+\tan^2\theta_x+\tan^2\theta_y}}\mbox{.}
\label{eqn:ceqn}
\end{equation}
Estimates for the angles of incidence $\theta_x$ and $\theta_y$ follow from a 
source localisation with the unmodified mask pattern $m(x)$ in a first step.
The finite spatial resolution of the position sensitive proportional counter is
described by convolution with a Gaussian integration kernel as in equation 
\ref{eqn:resolution}:

\begin{equation}
r(x) = \frac{1}{\sqrt{2\pi}c}\exp\left(-\frac{x^2}{2c^2}\right)
\mbox{ with }c=1/\sqrt{8\cdot\ln 2}\mbox{ mm.}
\label{eqn:resolution}
\end{equation}

The standard deviation corresponds to a value for full width half maximum of 
$1\mbox{ mm}$. In contrast to $a$, $c$ depends only very weakly on the spectral 
distribution of the incident photons and is considered to be constant.
After convolution with $r(x)$, the mask pattern has been modified to 
$m''(x)=[r\odot m'](x)$. $m''(x)$ is the image of the mask pattern under ideal
statistics and the source reconstruction with $m''(x)$ should not display any 
deviation from the ideal behaviour. The correlation $[f\otimes m''](x)$ yields
a corrected value $\theta_x'$ for the angle of incidence.

\subsection{Correlation in Fourier Space}
Due to their high numerical complexity it is favourable to perform both the 
convolutions and the image localisation by correlation in the Fourier domain. 
The nomenclature is such that lower case letters denote functions in real 
space and the corresponding upper case letters their Fourier transforms:

\begin{equation}
M(k)=\mathcal{F}\left[m(x)\right]=\frac{1}{2\pi}\int m(x)\exp(-ikx)dx\mbox{.}
\label{eqn:fourierpair}
\end{equation}

Convolutions and correlations reduce by virtue of equation \ref{eqn:fconvolve}  to 
mere multiplications in Fourier space. The asterisk denotes complex conjugation:

\begin{equation}
\begin{array}{lcl}
\left[p\odot m\right](x)&=&\mathcal{F}^{-1}\left[P(k)M(k)\right]\mbox{,}\\
\left[f\otimes m''\right](x)&=&\mathcal{F}^{-1}\left[F(k)^*M''(k)\right]\mbox{.}
\end{array}
\label{eqn:fconvolve}
\end{equation}

Diagram \ref{dia:cc} summarises all steps. Starting from the
Fourier-transform of the mask pattern $M(k)=\mathcal{F}[m(x)]$, both 
convolutions and the correlation are carried out in Fourier-space by
determining the product $C(k) =F(k)^* R(k) P(k) M(k)$. Inverse transformation yields
the correlation function $c(x)=\mathcal{F}^{-1}[C(k)]$, from which 
the corrected angle of incidence $\theta_x'$ is derived:

\begin{equation}
\begin{CD}
m(x) @>{\odot p(x)}>> m'(x) @>{\odot r(x)}>> m''(x) @>{\otimes f(x)}>> c(x) \\
@V{\mathcal{F}}VV @. @. @AA\mathcal{F}^{-1}A\\
M(k) @>>{\cdot P(k)}> M'(k) @>>{\cdot R(k)}> M''(k) @>>{\cdot F(k)^*}> C(k)\mbox{.}
\end{CD}
\label{dia:cc}
\end{equation}

In comparision to the {\em WFC} camera on board the BeppoSAX satellite, 
the penetration effect is noticably more pronounced in the case of HETE-2. 
Evaluating the average attenuation length for a photon at the corner of the 
field of view projected onto the detector plane determines the angular deviation 
in the worst case. This value for {\em WFC} is $\sim 7$ times smaller compared to 
{\wxm}, because of the narrower field of view, the higher gas pressure, which 
results in a shorter average attenuation length and the larger distance 
between the mask and detector. Furthermore, being a genuine 2-dimensional 
camera, the aberration vector field of {\em WFC} is radial and 
does not have the complex angular dependence described by equation 
\ref{eqn:ceqn}.

\section{Results\label{cont:result}}

\subsection{Monte Carlo Simulations}
\subsubsection{Simulated Data}
In order to investigate to which extend the algorithm is capable of correcting
systematic deviations, a set of $2^{14}$ Monte Carlo images containing a large number
of photons ($n_\gamma=3000$) was generated, so that the statistical scatter in the
source reconstruction is small. The simulated burst positions were randomly distributed, 
the angles of incidence ranging from $-30^\circ$ to $+30^\circ$. The number of incident 
photons was corrected with a factor of $\sqrt{1+\tan^2\theta_x+\tan^2\theta_y}$ for the
purpose of coping for the decreasing projected area of the detector at increasing
angles of incidence. The spectral distribution of photons followed a power law with 
spectral index of $\Gamma=1.1$, which is considered to be shallow, but not untypical for
$\gamma$-ray bursts \cite{yoshida}. Because of the flatness of the spectrum,
the aberration effect due to penetrating photons is distinctive and easy
to study. Localisation of photons by the proportional counter was modelled by
employing detector response data taken prior to the launch.

\subsubsection{Imaging Aberrations}
\begin{figure}[htb]
\begin{center}
	\mbox{
  	\epsfysize5cm
	\epsffile{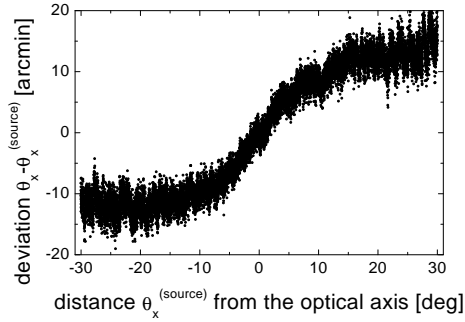}}
\end{center}
\caption{Deviation of the reconstructed event position $\theta_x$ from the
nominal position $\theta_x^{\scriptsize{(source)}}$ as a function of the
distance $\theta_x^{\scriptsize{(source)}}$ from the center of the field of
view following from a correlation of Monte Carlo generated images with
the unmodified mask pattern.}
\label{fig:pure}
\end{figure}

Figure \ref{fig:pure} shows the deviation 
$\theta_x-\theta_x^{\scriptsize{(source)}}$ of the uncorrected reconstructed
burst position from the nominal burst position as a function of angle of
incidence $\theta_x^{\scriptsize{(source)}}$. The aberration 
increases with increasing distance $\theta_x^{\scriptsize{(source)}}$ from the
center of the field of view. Each point corresponds to the localisation derived
from a Monte Carlo image.

Furthermore, one observes a widening of the scatter of 
$\theta_x-\theta_x^{\scriptsize{(source)}}$ with increasing
distance from the optical axis. This is due to the fact that the aberration, being
proportional to the function $a_{\|}(\theta_x,\theta_y)$ to first order, is 
concave for $\theta_x>0$ and convex for $\theta_x<0$. The magnitude of this 
curvature increases with increasing distance from the origin.

\subsubsection{Performance of the Algorithm\label{ana:mc}} 
\begin{figure}[htb]
\begin{center}
	\mbox{
  	\epsfysize5cm
	\epsffile{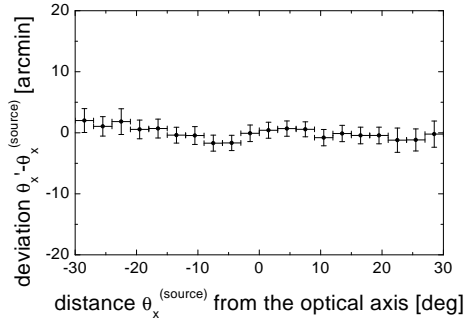}}
\end{center}
\caption{Deviation of the corrected event position $\theta_x'$ from the nominal source 
position $\theta_x^{\scriptsize{(source)}}$ as a function of angle of incidence
$\theta_x^{\scriptsize{(source)}}$ for the Monte Carlo generated dataset.}
\label{fig:mcdata_depend}
\end{figure}

In order to disentangle the contributions from statistical scatter
in the reconstruction from residual systematics, the dataset has been split into
$N=20$ intervals in $\theta_x^{\scriptsize (source)}$. For each interval $i$ the
mean value $\delta_i$ of $\theta_x'-\theta_x^{\scriptsize (source)}$ as well as 
the standard deviation $\sigma_i$ has been  
derived. Because the intervals are chosen such that they cover an angular 
range of just $3^\circ$, the scatter $\sigma_i$ can be assumed to be dominated 
by statistics. An increase in photon number $n_{\gamma}$ per image would have 
reduced the statistical scatter, according to 
$\sigma_i\sim 1/\sqrt{n_{\gamma}}$.\newline
From the values $\delta_i$ we derive the quantity 
$\delta=1/N\cdot\sqrt{\sum_i{\delta_i^2}}$, which is the mean quadratic 
deviation from zero. Further on, from $\sigma_i$ we determine the average 
statistical scatter $\sigma=1/N\cdot\sum_i\sigma_i$. From those two 
quantities, we define the spot size $\omega=\sqrt{\sigma^2+\delta^2}$.\newline
Figure \ref{fig:mcdata_depend} illustrates the performance of the applied
corrections: The mean quadratic deviation $\delta_i$ as well as the statistical
error $\sigma_i$ is plotted against the angle of incidence 
$\theta_x^{\scriptsize (source)}$. As the flatness indicates, $\delta_i$ is
consistent with zero, i.e. the imaging aberrations have been properly corrected.
Moreover, $\sigma_i$ has become independent from the angle of incidence
$\theta_x^{\scriptsize (source)}$, which indicates that the second order
influence of $\theta_y^{\scriptsize (source)}$ on the reconstruction has been
properly accounted for. Residual deviations are caused by detector 
nonlinearities, especially in the charge division method by which the position 
of an absorbed photon is determined.

\begin{figure}[htb]
\begin{center}
	\mbox{
  	\epsfysize6cm
	\epsffile{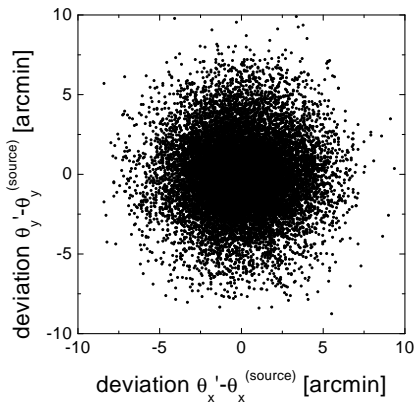}}
\end{center}
\caption{Scatter plot of all localisations 
$(\theta_x'-\theta_x^{\scriptsize{(source)}},
\theta_y'-\theta_y^{\scriptsize{(source)}})$ for the Monte Carlo set of images.}
\label{fig:mcdata_spot}
\end{figure}

The mean statistical scatter has been determined to $\sigma=1.53'\pm0.29'$ and 
the mean systematic deviation from zero to $\delta=1.00'\pm1.72'$. The 
algorithm thus fulfills the design requirements. 
Figure \ref{fig:mcdata_spot} shows the result of 
combining the reconstructions in $x$- and $y$-direction. The scatter plot is 
circularly symmetric and the width of the spot corresponds to 
$\omega=1.83'\pm0.97'$.

In the above analysis, the value for the average attenuation length $a$ has been
optimised is such a way that $\delta$ assumed a minimal value.
The smallest deviations resulted in the choice $a=3.7\mbox{ mm}$, 
which is in agreement with the theoretical value $a=4.0\mbox{ mm}$. The
theoretical value was derived by determining the flux weighted average of tabulated 
values for the photon attenuation length in xenon, where the photon spectrum has
been altered by the spectral photon absorption probability. 

\subsection{Astrometric Observation of {\em Scorpius X-1}}

\subsubsection{{\em Scorpius X-1} Data}
Data from the X-ray source {\em Scorpius X-1} has been taken for calibration purposes. 
Due to HETE-2's antisolar pointing, any source will drift through the field of view at 
ecliptic rate, i.e. $\sim1^\circ/\mbox{day}$. The orientation of the satellite was 
such that the {\wxm} coordinate system was rotated $\sim 10^\circ$ relative to the 
celestial $(\alpha,\delta)$ frame. {\em Scorpius X-1} was viewed at angles of incidence 
ranging from $\theta_x=-5^\circ$ to $\theta_x=+25^\circ$ and at $\theta_y=+5^\circ$ 
fixed. The spacecraft attitude was known to an accuracy surpassing $1'$.\newline

As pointed out in \cite{sco}, the spectrum of {\em Scorpius X-1}, being a low-mass 
X-ray binary, can be described by a multicolour blackbody spectrum. Due to the 
integration time of $\Delta t=2\mbox{ sec}$, short term spectral changes like 
quasi-periodic oscillations average out. Long term variabilities either happen on 
significantly longer time scales compared to the data taking, or they affect energies
outside HETE-2's sensitivity interval. Determining the average attenuation length by 
a photon flux weighted integral with tabulated values for the attenuation length yielded 
$a=2.5\mbox{ mm}$. Thus, the spectrum of {\em Scorpius X-1} is much softer compared to the 
previous case. The background subtracted images oconsisted roughly of $n_\gamma=3000$ photons.

\subsubsection{Performance of the Algorithm}
\begin{figure}[htb]
\begin{center}
	\mbox{
	\epsfysize5cm
	\epsffile{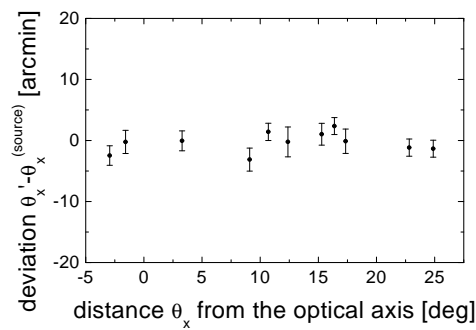}}
\end{center}
\caption{Deviation of the corrected event position $\theta_x'$ from the nominal source 
position $\theta_x^{\scriptsize{(source)}}$ as a function of angle of incidence
$\theta_x^{\scriptsize{(source)}}$ in the {\wxm}-coordinate frame for the
observation of {\em Scorpius X-1}.}
\label{fig:scodata_depend}
\end{figure}

The analysis \ref{ana:mc} is repeated for the {\em Scorpius X-1} dataset. For
each pointing $\delta_i$ and $\sigma_i$ have been derived. Figure 
\ref{fig:scodata_depend} illustrates the accuracy of the source localisation
algorithm. Again, the values $\delta_i$ with their errors $\sigma_i$ are shown
as a function of angle of incidence $\theta_x^{\scriptsize (source)}$.

The average quadratic deviation from zero has been determined to 
$\delta=1.59'\pm1.65'$ which is slightly larger than in the previous case, but
reflects inaccuracies in the satellite pointing. Furthermore, this value is
probably underestimated because of the incomplete sampling of the field of view. 
The statistical error is marginally smaller due to larger photon numbers and has 
been derived to be $\sigma=1.31'\pm0.51'$.

\begin{figure}[htb]
\begin{center}
	\mbox{
  	\epsfysize6cm
	\epsffile{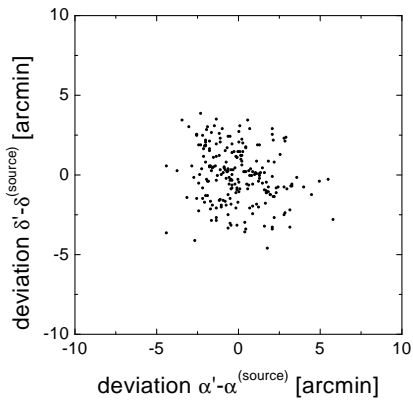}}
\end{center}
\caption{Scatter plot of all localisations 
$(\alpha'-\alpha^{\scriptsize{(source)}},\delta'-\delta^{\scriptsize{(source)}})$ 
in celestial coordinates. {\em Scorpius X-1} is located at the origin.}
\label{fig:scodata_spot}
\end{figure}

Figure \ref{fig:scodata_spot} shows the combined
localisation in $x$- and $y$-direction, converted into the celestial coordinate
frame. The radius of the spot is $\omega=2.06'\pm1.31'$.
The average attenuation length $a$ has been varied to yield the minimum value
for $\delta$. In the case of {\em Scorpius X-1}, we find the optimum value to be
$a=2.2\mbox{ mm}$, which corresponds well to the theoretically expected value
$a=2.5\mbox{ mm}$.

\section{Point Spread Function\label{cont:psf}}
\subsection{Autocorrelation Function}
In order to assess the extend to which angular resolving power is affected 
by the source localisation algorithm, the point spread function of the imaging 
system has been derived. For a coded mask camera like {\wxm}, the point spread 
function is equal to the autocorrelation $s(x)$ of the mask pattern $m''(x)$, 
i.e. $s(x)=[m''\otimes m''](x)$. To be exact, $s(x)$ describes the isotropised 
point spread function. Anisotropies provoke asymmetries in the central
correlation peak of the function $[m''\otimes m](x)$. Again, the determination
of $s(x)$ is carried out in Fourier-space. Diagram \ref{dia:psf} 
illustrates the derivation of $s(x)=\mathcal{F}^{-1}\left[|M''(k)|^2\right]$:

\begin{equation}
\begin{CD}
m(x) @>{\odot p(x)}>> m'(x) @>{\odot r(x)}>> m''(x) @>{\otimes m''(x)}>> s(x) \\
@V{\mathcal{F}}VV @. @. @AA\mathcal{F}^{-1}A\\
M(k) @>>{\cdot P(k)}> M'(k) @>>{\cdot R(k)}> M''(k) @>>{\vert\quad\vert^2}> S(k)\mbox{.}
\end{CD}
\label{dia:psf}
\end{equation}

\subsection{Variations in Angular Resolving Power}
The width $w_x$ of the autocorrelation function $s(x)$ has been determined by
fitting a Gaussian to the central peak. 
The average width $w$ of the point spread function has been defined as the 
geometric mean of its extend in $x$- and $y$-direction ($w^2=w_x\cdot w_y$), 
because the quantity of interest is the size of the spot a source is imaged onto.
The value of $s(x)$ at the origin corresponds to the correlation strength $v_x$.
Accordingly, the average correlation strength $v$ has been defined to be 
equal to the arithmetic mean of $v_x$ and $v_y$ because of its analogy to 
brightness.\newline

The angles of incidence have been restricted to
$\left|\theta^{\scriptsize{(source)}}\right|=\sqrt{\theta_x^2+\theta_y^2} <
20^\circ$, in order to ensure that the mask pattern is fully imaged onto the
detector. The source positions were arranged on a square lattice with mesh
size of $0.5^\circ$. The average attenuation length $a$ has been set to
\mbox{$3.7$ mm}, corresponding to a power law spectrum with spectral index
$\Gamma=1.1$.
\newline

\begin{figure}[htb]
\begin{center}
\begin{tabular}{cc}
	\mbox{
  	\epsfysize5cm
	\epsffile{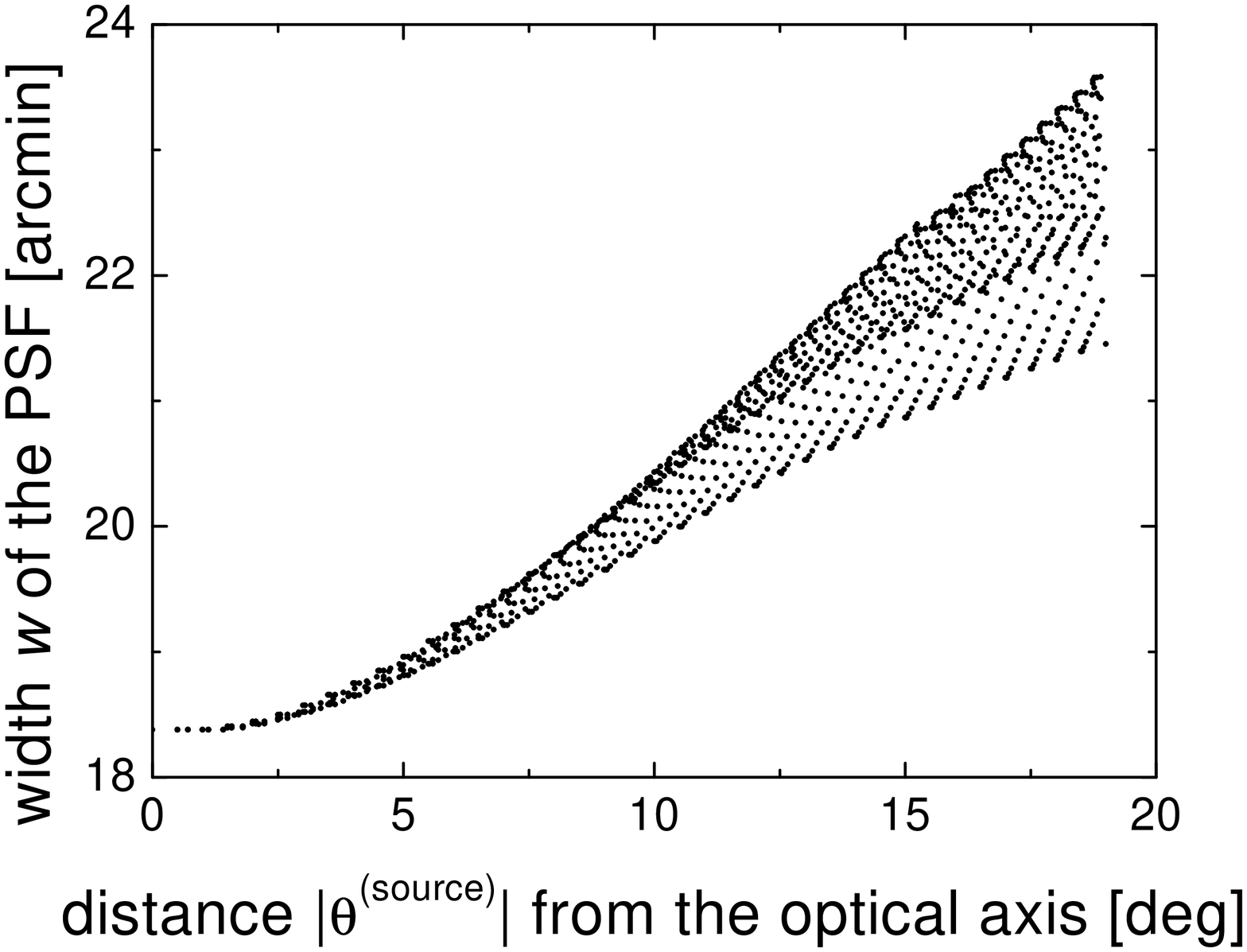}} &
	\mbox{
  	\epsfysize5cm
	\epsffile{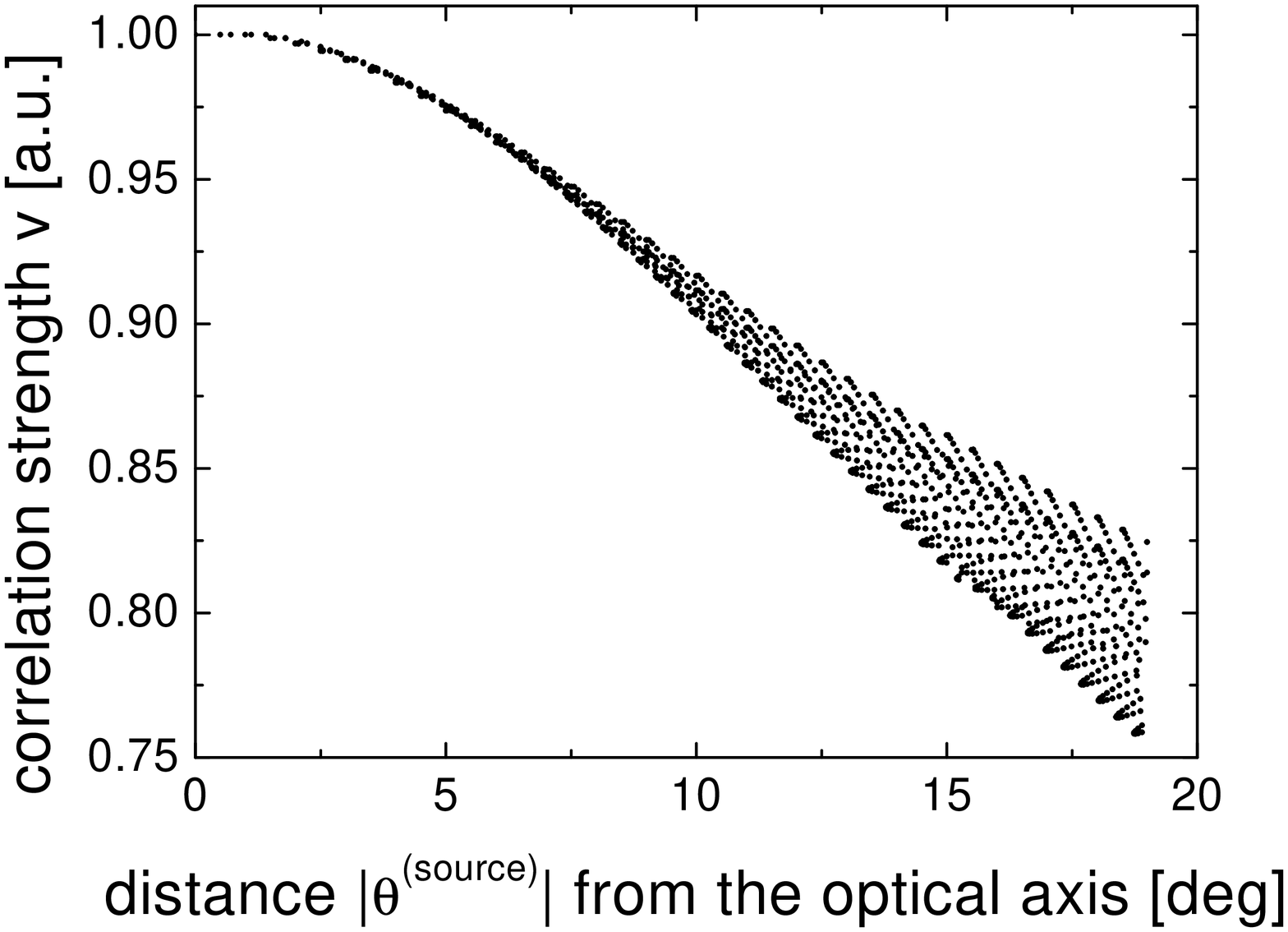}}
\end{tabular}
\end{center}
\caption{Average width $w$ in terms of FWHM (left) and correlation strength $v$ 
in arbitrary units (right) of the point spread function as a function of distance 
$\left|\theta^{\scriptsize (source)}\right|$ from the optical axis.}
\label{fig:psf}
\end{figure}

The left plot in Figure \ref{fig:psf} illustrates, that the width $w$ of the
autocorrelation function increases and how the correlation strength decreases
correspondingly at larger distances from the optical axis. At increasingly
larger distance from the optical axis, the functions $p(x)$ and $p(y)$ 
disperse the correlation peak. Because the area underneath the peak has to 
be conserved, the correlation strength $v$ drops accordingly. Thus, the angular
resolution power has decreased thus by $20\%$ for a burst at 
$\left|\theta^{\scriptsize (source)}\right|=20^\circ$ compared to a burst on the
optical axis.

The scatter in Figure \ref{fig:psf} is explained by the fact, that at fixed 
distance from the optical axis one obtains larger values of $w$ for a point 
situated on a diagonal than for a location on one of the axes, because in the 
latter case only one term is affected by the convolution with the functions 
$p(x)$ and $p(y)$ (formula \ref{eqn:penetrate}). An analogous argument applies 
to the dependence of $v$ on the angles of incidence. This effect is of the 
order of $8\%$ at $\left|\theta^{\scriptsize (source)}\right|=20^\circ$.

\section{Summary\label{cont:conclusion}}
This article describes an alternative correlation algorithm for $\gamma$-ray 
burst localisation with the {\wxm}-device on board the HETE-2 satellite. The
performance regarding source localisation accuracy has been validated using
Monte Carlo simulated data and astrometric observations of the X-ray
source {\em Scorpius X-1}.
\begin{itemize}
\item{The algorithm corrects systematic deviations in the imaging process by
convolving the mask pattern with suitable kernels. The convolution and the
source reconstruction are both carried out in Fourier space, which is favourable
with respect to time demand. Currently, the on board reconstruction algorithm
relies on a library of Monte Carlo generated images.}
\item{The Fourier algorithm is able to correct the systematic deviation caused
by photons penetrating into the depth of the detector with high accuracy
throughout the field of view (compare Table \ref{tab:result}). The residual 
mean quadratic deviation  is most likely due to nonlinearities in the detector 
response.}
\item{The influence of the convolutions on angular resolving
power have been investigated by determining the point spread function. One
observes a widening of the correlation peak by $\sim 30\%$ together with a drop
in correlation strength with increasing distance from the optical axis.} 
\item{The algorithm requires but a single, physically meaningful parameter: The
average attenuation length $a$. Optimised values for $a$ correspond to
theoretically expected values for a source with given spectral properties 
within $12 \%$ (compare Table \ref{tab:attenuate}). Theoretical values follow from 
formula \ref{eqn:averaging}:
\begin{equation}
a=\frac{\int\!\int\Phi(E) p(E,\vec{\theta}) a(E)d^2\theta dE}
{\int\!\int\Phi(E) p(E,\vec{\theta}) d^2\theta dE}\mbox{,}
\label{eqn:averaging}
\end{equation}
where $\Phi(E)$ is the photon spectrum, $p(E,\vec{\theta})$ the probability of
detecting a photon and $a(E)$ the attenuation length of a photon of energy
$E$. Because of the spectral diversity of {\grb}s, it is necessary to compute 
$a$ for each event separately. In an on board application the weighted sum over 
a parameterisation of the function $a(E)$ would replace integral 
\ref{eqn:averaging}.}
\end{itemize}

\begin{table}[htbb]
\begin{center}
\begin{tabular}{|l|cc|cc|}
\hline
 & \multicolumn{2}{c|}{Monte Carlo data} & \multicolumn{2}{c|}{{\em Scorpius X-1} data}\\
		& SD & FWHM & SD & FWHM \\
\hline
spot size 			$\omega$	& $1.83'\pm0.97'$ & $4.31'\pm2.28'$ & $2.06'\pm1.31'$ & $4.85'\pm3.08'$ \\
systematic error	$\delta$    & $1.00'\pm1.72'$ & $2.35'\pm4.05'$ & $1.59'\pm1.65'$ & $3.74'\pm3.88'$ \\
statistical error 	$\sigma$	& $1.53'\pm0.29'$ & $3.60'\pm0.68'$ & $1.31'\pm0.51'$ & $3.08'\pm1.20'$	\\
\hline
\end{tabular}
\end{center}
\caption{Summary of the localisation performance of the algorithm, stated in 
standard deviation (SD) and full width half maximum (FWHM), for both 
the Monte Carlo simulation (left column) and the astrometric observation 
of {\em Scorpius X-1} (right column).}
\label{tab:result}
\end{table}

\begin{table}[htb]
\begin{center}
\begin{tabular}{|l|c|c|}
\hline
				& Monte Carlo data 			& {\em Scorpius X-1} data\\
\hline
theoretical 	& $a=4.0\mbox{ mm}$  		& $a=2.5\mbox{ mm}$  \\
experimental 	& $a=3.7\mbox{ mm}$  		& $a=2.2\mbox{ mm}$  \\	
\hline
\end{tabular}
\end{center}
\caption{Comparison of theoretically expected with optimised values for the
spectrally averaged attenuation length $a$ for different photon distributions.
$a$ governs the correction for the imaging aberration due to 
photons penetrating into the depth of the detector.}
\label{tab:attenuate}
\end{table}

\section*{Acknowledgments}
B.M.S. wishes to thank Y. Shirasaki for help on the Monte Carlo generator. The support 
of the German National Merit Foundation is greatly appreciated. We thank all members 
of the HETE-2 team, who participated in the construction and operation of the satellite as 
well as the anonymous referee for valuable comments.


\end{document}